\documentclass[aps,pre,reprint,superscriptaddress,amsmath,amssymb,showpacs]{revtex4-1}
\usepackage{graphicx}
\usepackage{color}

\begin{document}

\title{Non-equilibrium fluctuations for linear diffusion dynamics}%
\author{Chulan Kwon}%
\email{ckwon@mju.ac.kr}
\affiliation{Department of Physics, Myongji University, Yongin, Gyeonggi-Do,
449-728, Republic of Korea}
\author{Jae Dong Noh}%
\email{jdnoh@uos.ac.kr}
\affiliation{Department of Physics, University of Seoul, Seoul 130-743,
Republic of Korea}
\affiliation{Department of Physics, Korea Institute for Advanced Study,
Seoul 130-722, Republic of Korea}
\author{Hyunggyu Park}%
\email{hgpark@kias.re.kr}
\affiliation{Department of Physics, Korea Institute for Advanced Study,
Seoul 130-722, Republic of Korea}

\date{\today}%

\begin{abstract}
We present the theoretical study on non-equilibrium (NEQ) fluctuations for diffusion dynamics in high dimensions driven
by a linear drift force. We consider a general situation in which NEQ is caused by two conditions:
(i) drift force not derivable from a potential function and (ii) diffusion matrix not proportional to the unit matrix,
implying non-identical and correlated multi-dimensional noise.
 The former is a well-known NEQ source and the latter can be realized in the presence of multiple heat reservoirs or multiple noise sources. We develop a statistical mechanical theory based on generalized thermodynamic quantities such as energy, work, and heat. The NEQ fluctuation theorems are reproduced successfully. We also
 find the time-dependent probability distribution function exactly as well as the NEQ work production distribution
 $P({\mathcal W})$ in terms of  solutions of nonlinear differential equations. In addition, we  compute low-order cumulants of the NEQ work production explicitly. In two dimensions, we carry out numerical simulations to check out our analytic results and also to get $P({\mathcal W})$. We find an interesting dynamic phase transition in the exponential tail shape of
 $P({\mathcal W})$, associated with a singularity found in solutions of the nonlinear differential equation.
 Finally, we discuss possible realizations in experiments.
\end{abstract}
\maketitle

\section{Introduction}
There have been great interests in non-equilibrium (NEQ) statistical
mechanics for last decades since the discovery of the fluctuation theorem
for entropy production. The first discovery was made on a deterministic
NEQ dynamics governed by the SLLOD
equation~\cite{evans,evans-searles1,gallavotti1}. Later on, the fluctuation
theorems of various types were found to be a universal feature for a wide class of NEQ systems, which are governed by both deterministic~\cite{gallavotti2,evans-searles2} and stochastic dynamics~\cite{crooks1,kurchan1,lebowitz,maes1,oono,sasa}. Jarzynski found an interesting relation between NEQ work and equilibrium free energy~\cite{jarzynski}, which was later proved to be a special case of Crooks fluctuation theorems~\cite{crooks2,kurchan1}. Since then, a number of stimulated studies have been published up to now regarding the fluctuation theorems and related phenomena~\cite{maes2,khkim,taniguchi,williams,komatsu,kurchan2,ge}.

The diffusion dynamics is distinguished from the Brownian dynamics. The
former has only position-like state variables that have even parity under the
time reversal, while the latter has pairs of position and momentum with even
and odd parities respectively. In this work, we consider the diffusion dynamics with
two important conditions which drive the system into a NEQ steady state (NESS):
(i) drift force not derivable from a potential function (ii) non-identical
and correlated noise. These two conditions can be realized only in
  high dimensions (not possible in one dimension).
  Unusual results were reported for the dynamics with the combination of these
  two conditions~\cite{ao1,yin,ao2}. In particular it was found that the zero mass limit and the over-damping limit
  are different in reducing the Kramers equation to the Fokker-Planck equation.

  Kwon, Ao and Thouless~\cite{kwon1}
  studied the diffusion dynamics with a linear drift force in high dimensions and found the probability distribution function (PDF) for the NESS exactly. They found that a circulating probability current can exist at the steady state, violating the detailed balance. One example is a non-zero torque generated in a nano heat engine in contact with two different heat reservoirs~\cite{filliger}. For a nonlinear drift force, it has recently been found via a perturbation theory that there exists an additional current, that is absent for the linear case, due to the combination of the force non-linearity and the  multi-dimensional noise correlation ~\cite{kwon2}. It moves the probability maximum away from the fixed point at which the force is zero. This novel current has the same origin with the noise-induced current, transporting drugs or molecules in biological systems, studied by Prost {\em et al.}~\cite{prost} and Doering {\em et al.}~\cite{doering}.

We revisit the diffusion dynamics with a linear drift force from the point of view of the NEQ fluctuation theorem, as one of a few analytically solvable cases far from equilibrium. We consider the Langevin equation
\begin{equation}
\dot{q}=f(q)+\xi~.
\label{langevin}
\end{equation}
where  $q=(q_1, q_2, \ldots, q_d)^T$ is a state vector
in $d$ dimensions with the superscript $T$ denoting the transpose of a given vector or matrix.
We restrict ourselves to the case in which $q$ has even parity under the time
reversal, i.e., there are no momentum-like variables.  $\xi =
(\xi_1,\xi_2,\cdots,\xi_d)^T $ is a white noise vector with zero mean satisfying
$\langle \xi(t)\xi^T(t')\rangle=2D\delta(t-t')$ where $\langle\cdots\rangle$ is the noise average.
$D$ is defined as a
$d\times d$ diffusion matrix that is symmetric, positive definite, and $q$-independent. The first NEQ
condition (i) is given by the drift force $f\neq -\nabla \Phi(q)$ where $\Phi$ is a scalar
function of $q$. The condition (ii) leads to the case where the diffusion
matrix is not proportional to the unit matrix, $D\not\propto  I$, in contrast to the conventional
thermal noise with $D\propto (k_B T) I$.

This Langevin equation describes a
general stochastic system far from equilibrium without conventional energy or temperature. So we first define
{\em generalized} thermodynamic quantities; energy, work, and heat properly.
With these definitions, we successfully reproduce the NEQ fluctuation theorems.
We can also get analytic expressions for many interesting quantities such as
the time-dependent PDF $P(q,t)$, the two-time correlation functions, and
the NEQ work production distribution $P({\mathcal W})$.

The Jarzynski equality can be shown directly and the cumulants for the NEQ work production ${\mathcal W}$ are calculated
explicitly up to the second order. More interestingly, we find the exponential tail shape of $P({\mathcal W})$
with a power-law prefactor, which undergoes a dynamic phase transition as the time increases. This
phase transition turns out to be associated with a singularity of solutions of a nonlinear differential equation (NLDE).
We solve this NLDE numerically to reveal the details of the exponential tail shape of $P({\mathcal W})$ and also its
dynamic phase transition.

This paper is organized as follows.
In Sec.~II, the generalized thermodynamic quantities are defined with the corresponding
fluctuation theorems. In Sec.~III, we obtain $P(q,t)$ and the two-time correlation functions
exactly. In Sec.~IV, we derive the analytic expression for the generating function of $P({\mathcal W})$ in terms of
solutions of the NLDE. In Sec.~V,
we calculate the cumulants of the work production.  In
Sec.~VI, we take the two-dimensional diffusion
dynamics as a simple example and calculate the generating function by solving the NLDE numerically.
The dynamic phase transition of the tail shape of $P({\mathcal W})$ is discussed.
We also present the results from the direct numerical integration of the Langevin equation, which agree with the analytical results. Finally, in Sec.~VII, we summarize our results and discuss novel features found for the linear diffusion dynamics in high dimensions and the possibility of realization in experiments.

\section{Fluctuation theorems for the diffusion dynamics in high dimensions}
\label{FT}

We consider the Fokker-Planck equation for the diffusion dynamics associated with the Langevin equation in Eq.~(\ref{langevin}),
\begin{equation}
\frac{\partial P(q,t)}{\partial t}=\nabla\cdot \left(-f(q)+D\cdot\nabla\right)P(q,t)~,\label{fokker-planck}
\end{equation}
where we use the dot notation for the product of
a vector and a matrix, or between vectors, but not between matrices. For example,
we write $q^T\cdot AB\cdot q$ where $A$, $B$ are matrices.

Writing the steady state solution as $P_{st}\propto
e^{-\Phi_{st}(q)}$, we define equilibrium as the steady state satisfying the detailed balance such that
\begin{equation}
\Pi(q',t';q,t)e^{-\Phi_{st}(q)}=\Pi(q,t;q',t')e^{-\Phi_{st}(q')},
\end{equation}
where $\Pi(q',t';q,t)$ is the conditional probability for the transition from state $q$ at time t to state $q'$ at time $t'$.

We can show that the necessary and sufficient condition for the detailed balance reads as
\begin{equation}
f(q)=-D\cdot\nabla\Phi_{st}(q),
\label{force}
\end{equation}
i.e., the vanishing probability current $j=(f-D\cdot\nabla)P_{st}=0$ at the steady state.
Defining the force matrix $F$ as
\begin{equation}
F_{\alpha\beta} = -\nabla_\beta f_\alpha,
\end{equation}
with $\nabla_\beta\equiv \frac{\partial}{\partial q_\beta}$, we can rewrite the detailed balance condition as
\begin{equation}
FD=(FD)^T=DF^{T}~,
\label{detailed-balance}
\end{equation}
where $\nabla_\alpha\nabla_\beta\Phi_{st}=\nabla_\beta\nabla_\alpha\Phi_{st}$ is used.

The detailed balance condition is
always satisfied in one dimension. In higher dimensions,
however, this  does not hold in general due to two possible sources.
One is the asymmetry of $F$ ($F\neq F^{T}$), which happens when $f$ is not derivable from a scalar
potential. Another comes from the diffusion matrix which is not proportional to the unit matrix,
so the noises are not identical in components ($D_{\alpha\alpha}\neq D_{\beta\beta}$ for $\alpha\neq \beta$)
and also may be correlated ($D_{\alpha\beta}\neq 0$).

We note that the detailed balance condition is satisfied (subsequently, equilibrium can be achieved)
not only with symmetric $F$ and $D\propto I$, but also with the specific combination of general $F$
and $D\not\propto I$ satisfying Eq.~(\ref{detailed-balance}). In equilibrium, the steady-state distribution
should be given by the Boltzmann distribution, so $\Phi_{st}$ can be interpreted as {\em energy}
(we set $k_B T\equiv 1$ for convenience). Then it is natural to define a generalized {\em force} as
the negative derivative of the energy function, i.e.~$D^{-1}\cdot f=-\nabla \Phi_{st}$  from Eq.~(\ref{force}),
especially for $D\not\propto I$.

In general, the detailed balance condition is not satisfied and the system is driven into a NESS.
In this case, Eq.~(\ref{force}) should be modified as
\begin{equation}
D^{-1}\cdot f=-\nabla\Phi(q)+g(q)~,
\label{Df}
\end{equation}
with nonzero $g(q)$ which can not be derivable from a scalar function ($g(q)\neq -\nabla \Psi(q)$).
We interpret $g$ as a generalized NEQ driving force. However, there is no unique way of
defining the NEQ driving force $g$ as well as the energy-like function $\Phi$.
As the NESS needs not be governed by the Boltzmann distribution,
the energy function $\Phi(q)$ does not have to be the same as $\Phi_{st}(q)$ in general.
In fact, one may choose an arbitrary $\Phi(q)$ and $g(q)$ accordingly in the system
described only by the stochastic equations like
Eqs.~(\ref{langevin}) and (\ref{fokker-planck}). We will come back to this issue later.

 Following the path integral formalism of Onsager and Machlup~\cite{onsager}, the conditional probability that the system
 evolves along a path $q(\tau)$ for $0\le\tau\le t$ starting from an initial state $q(0)$ is given by
\begin{equation}
\Pi[q(\tau)]_0^t\propto e^{ -\frac{1}{4}\int_{0}^{t}d\tau\left(\dot{q}(\tau)-f(q(\tau))\right)^{T}\cdot D^{-1}\cdot
\left(\dot{q}(\tau)-f(q(\tau))\right) }~.
\label{path-integral}
\end{equation}
The time-reverse path is given by $\bar{q}(\tau)=q(t-\tau)$ with the initial state $\bar{q}(0)=q(t)$.
Then the ratio of the conditional probability for the forward path $q(\tau)$ to that for the reverse path ${\bar q}(\tau)$
is found as
\begin{eqnarray}
\frac{\Pi[q(\tau)]_0^t}{\Pi[\bar{q}(\tau)]_0^t}&=&e^{\int_{0}^{t}d\tau \dot{q}^{T}\cdot D^{-1}\cdot f}
=e^{-\int_{0}^{t}d\tau\dot{q}^{T}\cdot(\nabla\Phi- g)}\nonumber\\
&=&e^{-\Delta\Phi+{\mathcal W}[q]}=e^{-{\mathcal Q}[q]}~,\label{detailed-fluctuation-relation}
\end{eqnarray}
where $\Delta \Phi=\Phi(q(t))-\Phi(q(0))$ is the energy difference between at the final and initial time. \
We interpret ${\mathcal W}[q]$ as the {\em work} production done by the NEQ force $g$ along the path $q(\tau)$,
\begin{equation}
{\mathcal W}[q]=\int_{0}^{t}d\tau \left(\dot{q}^{T}\cdot g+\frac{\partial\Phi}{\partial \tau}\right)~.
\label{work}
\end{equation}
The second term appears only when $\Phi$ has an explicit time dependence, which is the case of the Jarzynski type work~\cite{jarzynski}.

Accordingly we define ${\mathcal Q}\equiv\Delta\Phi-{\mathcal W}$ and interpret it as the {\em heat}
transferred into the system from the reservoir along the path. We can rewrite
\begin{equation}
{\mathcal Q}[q]=\int_{0}^{t}d\tau\left(-\dot{q}^{T}\cdot D^{-1}\cdot \dot{q}+\dot{q}^{T}\cdot \zeta\right)
\end{equation}
where $\zeta=D^{-1}\cdot(\dot{q}- f)=D^{-1}\cdot\xi$.
This definition of heat can be understood in terms of the corresponding Brownian dynamics.
The diffusion dynamics can be regarded as the Brownian dynamics in the overdamped limit (or zero inertia limit).
In this case, the corresponding Brownian dynamics is given by
\begin{equation}
m\ddot{q}=-\gamma\cdot \dot{q}+D^{-1}\cdot f+\zeta~,
\end{equation}
where the friction matrix $\gamma=D^{-1}$ and the diffusion matrix for the noise $\zeta=D^{-1}\xi$
becomes $D_\zeta=D^{-1}DD^{-1}=D^{-1}$. Therefore, the generalized Einstein relation is satisfied
$(D_\zeta=k_B T \gamma$ with our presetting of $k_B T=1$).
As noticed, $D^{-1}\cdot f$ plays the role of force. From the
point of view of the Brownian dynamics, $\zeta=D^{-1}\cdot\xi$ is a random
diffusive force exerted by noise. Then ${\mathcal Q}$ is the sum of dissipative work due to the random collisions and the diffusive work due to noise, known as the Onsager heat. Therefore the interpretation of ${\mathcal Q}$ as the heat production is reasonable.

Up to now, we introduced the appropriate definitions of generalized energy, work, and heat for an arbitrary diffusion
dynamics without conventional thermodynamic quantities. Note that the work and the heat production depend on the path, i.e.,
functionals of the path, while the energy is a state function and independent of the path.  One may restore the temperature by scaling the diffusion matrix as $D\to (k_B T) D$, then $\Phi_{st}$ scales as $\Phi_{st}/(k_B T)$. $k_B T$ can parametrize the overall strength of noise and can be interpreted as an effective temperature, if necessary. There are other temperature-like parameters, the eigenvalues of $D$, which may be different each other. The presence of multi and heterogeneous temperatures might also cause NEQ, though not always.

Eq.~(\ref{detailed-fluctuation-relation}) is the fundamental equation from which various fluctuation theorems can be derived. It replaces the detailed balance relation for equilibrium, so is referred to as the detailed fluctuation relation or the generalized detailed balance relation for NEQ. If we choose an initial PDF $\propto e^{-\Phi(q)}$ (Boltzmann distribution with the energy function $\Phi(q)$), we can easily show the Crooks fluctuation theorem~\cite{crooks1}:
\begin{equation}
\langle {\cal A}[q]\rangle_{F}=\langle \hat{\cal A}[\bar{q}]e^{\hat{\mathcal W}[\bar{q}]}\rangle_{R}~e^{-\Delta {\mathcal F}}~,
\label{FT_relation}
\end{equation}
where ${\cal A}[q]$ is any functional of the path $q(\tau)$ and $\hat{\cal A}[\bar{q}]\equiv{\cal A}[q]$.
The free energy ${\mathcal F}$ is defined as $e^{-{\mathcal F}}\equiv \int dq e^{-\Phi}$
and Eq.~(\ref{work}) guarantees that $\hat{\mathcal W}[\bar{q}]=-{\mathcal W}[\bar{q}]$.
The subscripts $F$ and $R$ denote the averages along the forward $q(\tau)$ and the reverse $\bar{q}(\tau)$ path respectively.
The free energy difference $\Delta {\mathcal F}={\mathcal F}(t)-{\mathcal F}(0)$ does not vanish only when $\Phi$ has an explicit time dependence ($\frac{\partial\Phi}{\partial t}\neq 0$). In this paper, we assume no explicit time dependence in $f$ and $D$, so we always find $\Delta {\mathcal F}=0$.

\section{Transient state far from steady state for linear diffusion dynamics}

From now on, we focus on the case of a linear drift force,
\begin{equation}
f=-F\cdot q,
\end{equation}
which can be analytically tractable.
The force matrix $F$ is constant of state $q$ and time $t$.
The exact steady state probability distribution $P_{st}\propto e^{-\Phi_{st}}$
was found by Kwon {\em et al.}~\cite{kwon1}
with
\begin{equation}
\Phi_{st}=\frac{1}{2}q^{T}\cdot U\cdot q~, \label{steady-state}
\end{equation}
where the symmetric matrix $U=U^T$ is given by
\begin{equation}
U=(D+Q)^{-1} F
\label{U}
\end{equation}
with the anti-symmetric matrix $Q=-Q^T$ satisfying
\begin{equation}
FQ+QF^{T}=FD-DF^{T}~. \label{anti-symmetric-matrix}
\end{equation}
The equilibrium detailed balance condition, Eq.~(\ref{detailed-balance}), yields $Q=0$ and $U=D^{-1} F$, which
satisfies Eq.~(\ref{force}) as expected.  Thus the existence of nonzero $Q$ implies the breaking of the detailed balance
and results in the non-zero steady state current as $j_{st}= -(Q\cdot\nabla\Phi_{st})P_{st}$.
The general solution for $Q$ can be found
in a series form
by using the Jordan transformation for asymmetric $F$ or in an integral form in the frequency space~\cite{kwon1,kwon2}.

The time-dependent PDF solution for the Fokker-Planck equation of Eq.~(\ref{fokker-planck}) can be formally found from the path integral, using Eq.~(\ref{path-integral}), as
\begin{equation}
P(q,t;l)=\int dq(0)P(q(0))\int D[q]e^{-\int_{0}^{t}d\tau \left(L(q,\dot{q})-l^{T}\cdot q\right)}~,
\label{path_1}
\end{equation}
where $P(q(0))$ is the initial PDF and the Lagrangian $L$  reads as
\begin{equation}
L(q,\dot{q})=\frac{1}{4}(\dot{q}+F\cdot q)^{T}\cdot D^{-1}\cdot (\dot{q}+F\cdot q)~.
\label{lagrangian}
\end{equation}
$\int D[q]\cdots$ denotes the integration over all paths reaching a fixed final state $q(t)$ at time $t$, starting
from an initial state $q(0)$, with proper normalizations.
The source field $l(\tau)$ is introduced for later use to generate time correlation functions and
moments of $q$.

We calculate explicitly the time-dependent PDF $P(q,t)=P(q,t;l=0)$ with the initial Gaussian PDF
of $P(q(0))\propto e^{-\frac{1}{2} q^T(0)\cdot A(0) \cdot q(0)}$ with a symmetric $A(0)$.
The detailed calculation steps are
given in Appendix~\ref{path-integration}. From  Eq.~(\ref{time-dep-prob2}), we get
\begin{equation}
P(q,t)=|\det(2\pi A^{-1}(t))|^{-1/2}e^{-\frac{1}{2}q^{T}\cdot A(t)\cdot q}~,
\label{kernel}
\end{equation}
where the symmetric matrix
$A^{-1}(t)$ is given from Eq.~(\ref{time-dep-variance1}) as
\begin{eqnarray}
A^{-1}(t)&=&U^{-1}+e^{-Ft}(A^{-1}(0)-U^{-1})e^{-F^{T}t}~.
\label{time-dep-variance2}
\end{eqnarray}
As $\lim_{t\rightarrow\infty} A^{-1} (t)=U^{-1}$ for positive definite $F$,
the steady-state solution of Eq.~(\ref{steady-state}) is recovered.

The differential equation for $A^{-1}$ can be derived from the recursion relation in Eq.~(\ref{recursion1})
as~\cite{FP-derivation}
\begin{equation}
\frac{d{A}^{-1}}{dt}=2D-FA^{-1}-A^{-1}F^T~.
\label{diff-eq1}
\end{equation}
One can easily show that $A^{-1}(t)$ in Eq.~(\ref{time-dep-variance2}) is the solution of this differential equation.

Now we define the generating functional for time-correlation functions and cumulants of $q$ as
\begin{equation}
Z[l(\tau)]=\int dq P(q,t;l(\tau))=e^{\frac{1}{2}\int d\tau\int d\tau' l^{T}(\tau)\cdot\Gamma(\tau, \tau')\cdot l(\tau')}
\end{equation}
In Appendix~\ref{generating-functional}, we find from Eqs.~(\ref{diagonal}) and (\ref{off-diagonal})
\begin{equation}
\Gamma(\tau, \tau')=
\left\{
  \begin{array}{ll}
    e^{-(\tau-\tau')F}A^{-1}(\tau'), & \hbox{$\tau>\tau'$,} \\
    A^{-1}(\tau)e^{- (\tau'-\tau) F^T}, & \hbox{$\tau'>\tau$,}
  \end{array}
\right.
\label{correlation}
\end{equation}
where $\Gamma(\tau,\tau^\prime)=\Gamma^T(\tau^\prime,\tau)$.
Then we can compute the time average of any functional ${\mathcal A}$ of path $q(\tau)$:
\begin{equation}
\langle {\mathcal A}[ q(\tau)]\rangle = \left.{\mathcal A}\left[\frac{\delta}{\delta l(\tau)}\right]Z[l]\right|_{l\to 0}~,
\label{average}
\end{equation}
with $Z[0]=1$. For example, we get the two-time correlation function as
\begin{eqnarray}
\langle q_\alpha(\tau) q_\beta(\tau^\prime)\rangle &=&\left.\frac{\delta}{\delta l_\alpha(\tau)}
\frac{\delta}{\delta l_\beta(\tau^\prime)}Z[l]\right|_{l\to 0}\nonumber\\
&=& \Gamma_{\alpha\beta} (\tau, \tau^\prime)=\Gamma_{\beta\alpha} (\tau^\prime, \tau)~.
\label{two-point}
\end{eqnarray}

\section{Non-equilibrium work production}
The work production by the NEQ force $g$ along the path $q(\tau)$ is given by Eq.~(\ref{work})
with $\frac{\partial\Phi}{\partial \tau}=0$. As discussed in Sec.~II, we have some arbitrariness
in choosing the energy functional $\Phi(q)$, thus also the NEQ force $g(q)$ in Eq.~(\ref{Df}).

In the case of the linear drift force  with $D^{-1}\cdot f =-D^{-1}F\cdot q$,
Eq.~(\ref{Df}) becomes
\begin{equation}
\left(D^{-1}F\right)_{\alpha\beta}=\nabla_\beta \nabla_\alpha \Phi
-\nabla_\beta g_\alpha~.
\end{equation}
If $D^{-1}F$ is symmetric, the detailed balance condition, Eq.~(\ref{detailed-balance}) is satisfied
and one may choose $\Phi=\frac{1}{2} q^T\cdot (D^{-1}F)\cdot q$ with $g=0$. So we get
no NEQ work production with this choice of the energy function, as expected.

When $D^{-1}F$ is not symmetric, we must have a nonzero NEQ force $g$. The energy function
$\Phi(q)$  can be written in general as
\begin{equation}
\Phi(q)=\frac{1}{2} q^T\cdot G_s\cdot q~,
\end{equation}
with a symmetric matrix $G_s=G_s^T$ which is a part of $D^{-1}F$. Then we can
divide $D^{-1}F$ into the symmetric part $G_s$ and the remainder  $G_a$:
\begin{equation}
D^{-1}F=G_s+G_a~,
\end{equation}
and the NEQ driving force is given as
\begin{equation}
g(q)=-G_a\cdot q~.
\end{equation}

There is no unique way to determine $G_s$ or $G_a$ out of $D^{-1}F$. One natural possible choice
is to enforce $G_a$ anti-symmetric ($G_a=-G_a^T$), such as
\begin{eqnarray}
G_s&=&\frac{1}{2}(D^{-1}F+F^TD^{-1})\equiv\bar{G_s}~, \nonumber\\
G_a&=&\frac{1}{2}(D^{-1}F-F^TD^{-1})\equiv\bar{G_a}~,
\label{antisymmetric-gauge}
\end{eqnarray}
which will be called the {\em anti-symmetric} (AS) choice.

Another interesting choice, called as the {\em steady-state} (SS) choice, is
\begin{equation}
G_s=U~,~~G_a=D^{-1}QU~.
\label{invariant-gauge}
\end{equation}
If we take an initial Boltzmann distribution with this energy function, the system stays in the NESS from the beginning.

In general, one can choose
\begin{eqnarray}
G_s &=& \bar{G}_s + \delta G_s~, \nonumber \\
G_a &=& \bar{G}_a - \delta G_s~,
\end{eqnarray}
with an arbitrary symmetric matrix $\delta G_s$.
From Eq.~(\ref{work}), the NEQ work production during time $t$ is given by
\begin{equation}
{\mathcal W}[q]=-\int_{0}^{t} d\tau\dot{q}^{T}\cdot G_a \cdot q
=\bar{\mathcal W}+\delta{\mathcal W}~,
\label{work-production}
\end{equation}
where $\bar{\mathcal W}$ is the NEQ work production in the AS choice
\begin{equation}
 \bar{\mathcal W}[q]= -\int_0^t d\tau \dot{q}^T \cdot \bar{G}_a \cdot q~.
 \label{barW}
 \end{equation}
Contribution $\delta {\mathcal W}$ from the additive symmetric matrix $\delta G_s$ is
given by
\begin{eqnarray}
\delta {\mathcal W} &=& \int_0^t d\tau \dot{q}^T \cdot \delta G_s \cdot q \nonumber\\
&=& \frac{1}{2} q^T(t)\cdot \delta G_s\cdot q(t)-\frac{1}{2}
q^T(0)\cdot \delta G_s\cdot q(0)~, \label{variation_pdf}
\end{eqnarray}
which comes only from boundaries and also exactly compensates the additional energy term due to $\delta G_s$
in the energy function $\Phi$.

It is important to note that
the heat ${\mathcal Q}=\Delta\Phi-{\mathcal W}$ is independent of the choice of $\delta G_s$,
in contrast to the NEQ work ${\mathcal W}$. In the long-time limit, $\delta {\mathcal W}$
becomes negligible as the NEQ work usually increases incessantly in time. Thus the main contribution
to the NEQ work production in the steady state comes from the purely anti-symmetric part $\bar{\mathcal W}$.

We now consider the generating function ${\mathcal G}(\lambda)$ for the PDF of the NEQ work production $P({\mathcal W})$ as
\begin{eqnarray}
\label{G_lambda}
{\mathcal G}(\lambda)&=&\langle e^{-\lambda{\mathcal W}}\rangle \nonumber\\
&=&\int dq(t)  dq(0) P(q(0))
\int D[q] e^{-\int_0^t d\tau L(q,\dot{q})-\lambda{\mathcal W}[q]}~, \nonumber\\
&=&\int d{\mathcal W} P({\mathcal W})e^{-\lambda{\mathcal W}},
\end{eqnarray}
with the initial equilibrium Boltzmann distribution $P(q(0))\propto e^{-\frac{1}{2} q^T(0)\cdot G_s \cdot q(0)}$.
The PDF of the work production $P({\mathcal W})$ can be obtained formally by
 \begin{equation}
P({\mathcal W})=\int\frac{d\lambda}{2\pi}e^{i\lambda {\mathcal W}+\ln{\mathcal G}(i\lambda)}~.
\label{PW}
\end{equation}

We calculate ${\mathcal G}(\lambda)$ in a similar way in which the path integral is computed in Appendix~\ref{path-integration}
for the time-dependent PDF, $P(q,t)$. As ${\mathcal W}$ is quadratic in $q$,
the integral in Eq.~(\ref{G_lambda}) is basically the same as the integral in Eq.~(\ref{path_1}) except for
the final integral over $q(t)$ with the {\em modified} Lagrangian as
\begin{equation}
L=\frac{1}{4}(\dot{q}+\tilde{F}\cdot q)^{T}\cdot D^{-1}\cdot (\dot{q}+\tilde{F}\cdot q)+\frac{1}{2}q^T\cdot\Lambda\cdot q~,
\end{equation}
where
\begin{eqnarray}
\tilde{F}&=&F-2\lambda D\bar{G}_a\\
\Lambda &=&\frac{1}{2}\left(F^TD^{-1}F-\tilde{F}^TD^{-1}\tilde{F}\right)\\
&=& \lambda \left(F^T\bar{G}_a - \bar{G}_a F + 2\lambda ~\bar{G}_a D \bar{G}_a \right)\nonumber~.
\end{eqnarray}
The contribution from $e^{-\lambda\delta{\mathcal W}}$ only modifies the initial and final distribution according to Eq.~ (\ref{variation_pdf}).

Before going further, we briefly comment on the discrete-time representation of
the path integral and the work ${\mathcal W}$. In Appendix~\ref{path-integration}, we perform the path integral in the discrete-time representation.
Choice of the $q(\tau)$ value between the discrete time interval $\Delta t$ does not affect
the PDF at final time $t$ in the limit of $\Delta t\rightarrow 0$. However, it is well known that
one should choose the midpoint $q$ value for the definition of the work for the correct description~\cite{midpoint}.
So $\bar{\mathcal W}$ in Eq.~(\ref{barW}) should be written as
\begin{eqnarray}
\bar{\mathcal W}&=&-\frac{1}{2}\sum_{i=1}^{N}(q_{i}-q_{i-1})^T\cdot\bar{G}_a\cdot(q_{i}+q_{i-1}) \nonumber\\
&=&-\sum_{i=1}^{N}(q_{i}-q_{i-1})^T\cdot\bar{G}_a\cdot q_{i-1}~,\label{work-prepoint}
\end{eqnarray}
where $q_i=q(t_i)$ for $i=0,\ldots,N$ with $\Delta t=t/N$.
Note that it is first expressed in the mid-point representation but becomes identical to the so-called pre-point representation due to the anti-symmetricity of $\bar{G}_a$. In Appendix~\ref{path-integration}, all calculations
are done in the pre-point representation for convenience.

First, we perform the path integral of Eq.~(\ref{G_lambda}) without the final integral over $q(t)$.
The integration procedure is basically identical to the case for the time-dependent PDF calculation
in Appendix~\ref{path-integration} except for the different initial condition and the modified Lagrangian.
Let $\tilde{A}(t;\lambda)$ be the modified kernel for $A(t)$ of Eq.~(\ref{kernel}). In the discrete-time representation,
the recursion relation Eq.~(\ref{recursion1}) in Appendix~\ref{path-integration}  is modified as
\begin{equation}
\tilde{A}_i^{-1}=2\Delta t D+\tilde{V}\left(\tilde{A}_{i-1}+\Delta t
\Lambda\right)^{-1}\tilde{V}^T~,
\label{diff-eq2-0}
\end{equation}
with $\tilde{V} = 1-\Delta t \tilde{F}$.
Taking $\Delta t\to 0$ limit, the differential equation for $\tilde{A}^{-1}$ can be derived,
\begin{equation}
\frac{d{\tilde A}^{-1}}{dt}=2D-\tilde{F}\tilde{A}^{-1}-\tilde{A}^{-1}\tilde{F}^T -\tilde{A}^{-1}\Lambda \tilde{A}^{-1}~.
\label{diff-eq2}
\end{equation}
In contrast to Eq.~(\ref{diff-eq1}), this is a nonlinear differential matrix equation, which can not
be solved analytically in general. Using $d\tilde{A}^{-1}/dt=-\tilde{A}^{-1} (d\tilde{A}/dt)\tilde{A}^{-1}$,
we can rewrite this equation as
\begin{equation}
\frac{d\tilde{A}}{dt}=-2\tilde{A}D\tilde{A}+\tilde{A}\tilde{F}+\tilde{F}^T\tilde{A} +\Lambda~.
\label{diff-eq3}
\end{equation}
The initial condition is given as $\tilde{A}(0;\lambda)=\bar{G}^{s}+(1-\lambda)\delta G_s$,
where the $\lambda$-dependent term comes from $\delta{\mathcal W}$. We solve this equation numerically
for a specific case in Sec.~VI.

We obtain ${\mathcal G}(\lambda)$  after integrating over the final $q(t)=q_N$ in Eq.~(\ref{G_lambda}),
leading to
\begin{equation}
{\mathcal G}(\lambda)=\left|\frac{\det(\tilde{A}_N+\lambda\delta G_s)}{\det(\bar{G}_s+\delta G_s)}\right|^{-1/2}
\prod_{i=0}^{N-1}\left|\frac{\det(\tilde{A}_{i}+\Delta t\Lambda)}{\det(\tilde{A}_{i+1})}\right|^{-1/2}~.
\end{equation}
Using $\tilde{A}_{i+1}=\tilde{A}_i+(\Delta t) d\tilde{A}_i/dt$ for the denominator, we finally get
\begin{eqnarray}
\ln{\mathcal G}(\lambda)&=&-\frac{1}{2}\int_{0}^{t}d\tau \mbox{Tr} \left(\Lambda-
\frac{d{\tilde{A}}(\tau;\lambda)}{dt}\right) \tilde{A}^{-1}(\tau;\lambda)\nonumber\\
&&-\frac{1}{2}\ln\left(\frac{\det(\tilde{A}(t;\lambda)+\lambda\delta G_s)}{\det(\bar{G}_s+\delta G_s)}\right)\nonumber\\
&=& -\int_{0}^{t}d\tau \mbox{Tr}(\tilde{A}D-\tilde{F})\nonumber\\
&&-\frac{1}{2}\ln\left(\frac{\det(\tilde{A}(t;\lambda)+\lambda\delta G_s)}{\det(\bar{G}_s+\delta G_s)}\right)~.
\label{lnG}
\end{eqnarray}
It is not possible to perform the integral of Eq.(\ref{PW}) to find $P({\mathcal W})$ in a closed form. However,
the explicit form of ${\mathcal G}(\lambda)$ reveals many interesting properties of $P({\mathcal W})$. For example,
as $\ln{\mathcal G}$ is not quadratic in $\lambda$, $P({\mathcal W})$ is not Gaussian in general. Due to the
logarithmic boundary term, the divergence ${\mathcal G}(\lambda)$ may appear in $\lambda$, which determines the asymptotic behavior of a non-Gaussian tail
of $P({\mathcal W})$ for large $|{\mathcal W}|$. This will be investigated more in detail in Sec.~VI.

The fluctuation theorem yields
\begin{equation}
{\mathcal G}(\lambda)={\mathcal
G}(1-\lambda)
\label{FT_G}
\end{equation}
by substituting ${\mathcal A}$ with $e^{-\lambda {\mathcal W}}$ in Eq.~(\ref{FT_relation}).
The work ${\mathcal W}[\bar{q}]$ in the reverse path should be the same as $-{\mathcal W}[q]$
in the forward path, and the forward and reverse path are identical with the same initial conditions
with the same energy function. It seems not easy to prove Eq.~(\ref{FT_G})
for general $\lambda$, directly from Eq.~(\ref{lnG}).

However, we can prove ${\mathcal G}(1)=\langle e^{-{\mathcal W}}\rangle=1$ easily, which
corresponds to the Jarzynski equality for a time-dependent
potential. For $\lambda=1$, $\tilde{A}(0) =\bar{G}_s$, $\tilde{F} = D\bar{G}_s - D\bar{G}_a$, and $\Lambda = \bar{G}_s D \bar{G}_a - \bar{G}_a D \bar{G}_s$. We can show
the initial state is the fixed point of Eq.~(\ref{diff-eq2}) or (\ref{diff-eq3}), i.e., $\tilde{A}(t) = \bar{G}_s$. Then the logarithmic part vanishes in Eq.~(\ref{lnG}).
We can also see $\mbox{Tr}(\tilde{A}D-\tilde{F})=\mbox{Tr}D\bar{G}_a=0$.
Hence ${\mathcal G}(1)=1$.

\section{Cumulants of NEQ Work production}

In this section, we calculate the cumulants of the work production by using the two-time correlation function $\Gamma(\tau,\tau^\prime)$ in Eqs.~(\ref{correlation}) and (\ref{two-point}).
For simplicity, we take the AS choice where $G_a=\bar{G_a}$ , $\delta G_s=0$, $\delta{\mathcal W}=0$, and
${\mathcal W}=\bar{\mathcal W}$. However, in the long-time limit, all results are choice-independent.

First, consider the first cumulant of ${\mathcal W}$ in the discrete-time representation, Eq.~(\ref{work-prepoint}) as
\begin{eqnarray}
\langle {\mathcal W}\rangle
&=& -\sum_{i=1}^{N} \langle q^T_{i} \bar{G_a} q_{i-1}\rangle \nonumber\\
&=&\sum_{i=1}^{N}\mbox{Tr}\Gamma_{i,i-1}\bar{G_a}
\nonumber\\
&=&\sum_{i=1}^{N}\mbox{Tr}(1-\Delta t F)A^{-1}_{i-1}\bar{G_a}
\nonumber\\
&=&-\sum_{i=1}^{N}(\Delta t)\mbox{Tr} F A^{-1}_{i-1}\bar{G_a}\nonumber\\
&\to&-\int_0^t d\tau\mbox{Tr}A^{-1}(\tau)\bar{G_a}F~,
\end{eqnarray}
where we use $q_{i-1}^T\cdot\bar{G_a}\cdot q_{i-1}=0$, $\Gamma_{i,i-1}=(1-\Delta t F)A_{i-1}^{-1}$,
and Eq.~(\ref{two-point}).

In the long time limit, we can replace $A^{-1}$ by $U^{-1}$ and find
\begin{eqnarray}
\langle {\mathcal W}\rangle &\to & -t~\mbox{Tr}U^{-1}\bar{G_a} F \nonumber\\
&=& -t~\mbox{Tr}Q\bar{G_a}\nonumber\\
&=& t~\mbox{Tr}QF^TD^{-1}~,
\label{first-cumulant}
\end{eqnarray}
which should be choice-independent. Since the energy difference
$\langle\Delta\Phi\rangle$ is finite, $\langle {\mathcal W}\rangle\simeq -\langle {\mathcal Q}\rangle$ measures the entropy production piled up in the reservoir with the mean rate
\begin{equation}
\langle\sigma\rangle=t^{-1} \langle{\mathcal W}\rangle=\mbox{Tr}QF^TD^{-1}~.
\label{entropy-production-rate}
\end{equation}
It is expected to be positive from the fluctuation theorem, which implies that the second law of thermodynamics should hold for general NEQ phenomena with no thermodynamic origin.

The second cumulant of $\mathcal W$ can also be found as
\begin{eqnarray}
\lefteqn{\langle{\mathcal W}^2\rangle_c=\langle{\mathcal W}^2\rangle-\langle{\mathcal W}\rangle^2}\nonumber\\
&=&\sum_{i,j}\left\langle \left(q_{i}^T\cdot \bar{G}_a \cdot
q_{i-1}\right)\left(q_{j}^T\cdot \bar{G}_a\cdot q_{j-1}\right)\right\rangle\nonumber\\
&&-\left(\sum_{i}\left\langle q_{i}^T\cdot \bar{G}_a\cdot q_{i-1}\right\rangle\right)^2\nonumber\\
&=&\sum^\prime_{i,j}\mbox{Tr}\left(-\Gamma_{ij}\bar{G}_a\Gamma_{j-1i-1}
\bar{G}_a +\Gamma_{ij-1}\bar{G}_a\Gamma_{ji-1}\bar{G}_a \right)~,
\end{eqnarray}
where the summation $\sum^\prime_{i,j}=2\sum_{i>j} +\sum_{i=j}$ due to the symmetry between
$i$ and $j$.
 Using Eq.~(\ref{correlation}), we can write
\begin{eqnarray}
\lefteqn{\langle{\mathcal W}^2\rangle_c}\nonumber\\
&=&\sum_{i}\Bigg[ 2\sum_{j=1}^{i-1}\mbox{Tr}\left\{-e^{-F(t_{i}-t_{j})}A_j^{-1}\bar{G}_a A_{j-1}^{-1}e^{-F^T(t_{i-1}-t_{j-1})}\bar{G}_a\right.\nonumber\\
&&\left. +e^{-F(t_i-t_{j-1})}A_{j-1}^{-1}\bar{G}_a A_{j}^{-1}e^{-F^T(t_{i-1}-t_j)}\bar{G}_a\right\} \nonumber\\
&&+\mbox{Tr}\left(-A_{i}^{-1}\bar{G}_a A_{i-1}^{-1}\bar{G}_a+\right.\nonumber\\
&&+\left.e^{-F\Delta t}A_{i-1}^{-1}\bar{G}_ae^{-F\Delta t} A_{i-1}^{-1}\bar{G}_a\right)
\Bigg]~.
\end{eqnarray}
In $\Delta t\to 0$ limit, we get
\begin{eqnarray}
\lefteqn{\langle{\mathcal W}^2\rangle_c=}\nonumber\\
&&-2\mbox{Tr}\int_0^t d\tau e^{-F^T\tau}(F^T\bar{G_a}-\bar{G_a}F)Fe^{-F\tau}\nonumber\\
&&~~~~\times~\int_0^{\tau}d\tau' e^{F\tau'}A^{-1}(\tau')\bar{G}_a A^{-1}(\tau')e^{F^T\tau'}\nonumber\\
&&-2\mbox{Tr}\int_0^t d\tau e^{-F^T\tau}(F^T\bar{G_a}-\bar{G_a}F)e^{-F\tau}\nonumber\\
&&~~~~\times~\int_0^{\tau}d\tau' e^{F\tau'}\dot{A}^{-1}(\tau')\bar{G}_a A^{-1}(\tau')e^{F^T\tau'}\nonumber\\
&&-2\mbox{Tr}\int_0^td\tau \bar{G}_a FA^{-1}(\tau)\bar{G}_a A^{-1}(\tau) \nonumber\\
&&-\frac{1}{2}\mbox{Tr}\left(\bar{G}_a A^{-1}(t)\bar{G}_aA^{-1}(t)-\bar{G}_a A^{-1}(0)\bar{G}_aA^{-1}(0)
\right)~,
\label{second-cumulant}
\end{eqnarray}
where $\dot{A}^{-1}(\tau)$ is given by Eq.~(\ref{diff-eq1}).

Using the identity of Eq.~(\ref{integral}) found in the Appendix~\ref{path-integration}, one can perform
the above integral in principle. In this paper, rather than reporting the exact time-dependence of Eq.~(\ref{second-cumulant}), we compute the long-time behavior by keeping only the most dominant contributions,
\begin{eqnarray}
{\langle{\mathcal W}^2\rangle_c}&\to&-t~\mbox{Tr}\left(C+E\right)(F^T\bar{G_a}-\bar{G_a}F)\nonumber\\
&&-2t~\mbox{Tr}\bar{G_a}FU^{-1}\bar{G_a}U^{-1}\nonumber\\
&=&2t~\mbox{Tr}\left[\bar{G_a}F(E-C)\right]~
\label{work-variance}
\end{eqnarray}
where the matrix $C$ is anti-symmetric, defined by
\begin{equation}
C=U^{-1}\bar{G}_aU^{-1}~,
\label{C1}
\end{equation}
and the matrix $E$ is symmetric, determined by
\begin{equation}
FE+EF^T=FC-CF^T~.
\label{E1}
\end{equation}

Note that $\langle(t^{-1}{\mathcal W})^2\rangle_c\sim t^{-1}$. It implies that the PDF of the entropy production rate, $\sigma =t^{-1}\mathcal W$, shows a sharp distribution with the mean value $\langle\sigma\rangle$ found in Eq.~(\ref{entropy-production-rate}) and the variance of order $t^{-1}$. Assuming it as Gaussian,  $P(\sigma)\sim e^{-\frac{(\sigma-\langle \sigma\rangle)^2}{2\langle\sigma^2\rangle_c}}$, we obtain
\begin{equation}
\frac{P(\sigma)}{P(-\sigma)}=e^{ts\sigma}~,~~s=\frac{2\langle\sigma\rangle}{t\langle\sigma^2\rangle_c}~.
\label{gaussian}
\end{equation}
From Eq.~(\ref{work-variance}), $\langle\sigma^2\rangle_c\sim t^{-1}$. It qualitatively agrees with the fluctuation theorem for the entropy production. However, the fluctuation theorem predicts $s=1$, while it seems not equal to 1 if estimated by assuming the Gaussian distribution. This implies that the PDF for the work and entropy production is in general non-Gaussian and non-Gaussian tails make a significant contribution to the exact theorem.

More information on $P({\mathcal W})$ might come from higher cumulants in $W$. In principle, it can be done systematically by using Eq.~(\ref{average}), but it is too complicated to proceed further calculation in detail.

\section{Example: Diffusion in two dimensions}

Now, we take an example of a two dimensional diffusive motion, $q=(x,y)^T$ for more explicit calculations.
Consider
\begin{equation}
F=\left(
\begin{array}{cc}
k_1 & \kappa_1\\
\kappa_2 & k_2
\end{array}\right)~,~~
D=\left(
\begin{array}{cc}
\alpha & \epsilon\\
\epsilon & \gamma
\end{array}\right)~.
\end{equation}
By the orthogonal coordinate transformation, $D$ can be diagonalized and the calculation goes simpler. However, we keep the present form of $D$ in order to examine the effect of noise correlations. If $FD-DF^T\neq 0$,  we have a nonzero anti-symmetric matrix $Q$, a measure for NEQ, which can be obtained easily from Eq.~(\ref{anti-symmetric-matrix}),
\begin{equation}
Q=\left(
\begin{array}{cc}
0 & \hat{q}\\
-\hat{q} & 0
\end{array}\right)~,
~~\hat{q}=\frac{\epsilon(k_1-k_2)+\gamma\kappa_1-\alpha\kappa_2}{k_1+k_2}~.
\end{equation}

The system goes to equilibrium for $\hat{q}=0$. The conventional Gibbs-Boltzmann (GB) type equilibrium is a trivial case where the force is conservative ($\kappa_1=\kappa_2$), and the noises are identical and independent ($\alpha=\gamma$, $\epsilon=0$). The equilibrium PDF is given as $e^{-\alpha^{-1}{\mathcal E}(x,y)}$, and ${\mathcal E}(x,y)=\frac{1}{2}(k_1 x^2+2\kappa_1 xy+k_2 y^2)$. There are also non-trivial equilibria possible even for a non-conservative force ($\kappa_1\neq\kappa_2$) and non-identical/correlated noises ($\alpha\neq\gamma$, $\epsilon\neq 0$) as long as $\hat{q}=0$. In this case, the equilibrium PDF is given as $e^{-\frac{1}{2}q^T\cdot D^{-1}F\cdot q}$. There seems no fundamental difference among many equilibria from the view point of our paper in the sense that they are all preserving the detailed balance.

Now we consider the NESS for $\hat{q}\neq 0$. From Eq.~(\ref{entropy-production-rate}), the entropy production rate
in the long time limit is given as
\begin{equation}
\langle\sigma\rangle=t^{-1}\langle\mathcal W\rangle=\frac{\mbox{Tr}F}{\det D}\hat{q}^2~,
\end{equation}
where $\det D=\alpha\gamma-\epsilon^2$ and $\mbox{Tr} F=k_1+k_2$.

The second cumulant of the work can be obtained from Eq.~(\ref{work-variance}). The matrices $C$ and $E$ are found
from Eqs.~(\ref{C1}) and (\ref{E1}) as
\begin{equation}
C=\left(\begin{array}{cc}
0&c\\
-c&0
\end{array}\right)~,~~
E=\frac{c}{\mbox{Tr} F}
\left(
\begin{array}{cc}
-2\kappa_1 & k_1-k_2\\
k_1-k_2 & 2\kappa_2
\end{array}\right)~,
\end{equation}
with $c=\frac{\det(D+Q){\rm Tr}F}{2\det D\det F}\hat{q}$.
Then, $\langle \sigma^2\rangle$ is found as
\begin{eqnarray}
\lefteqn{\langle\sigma^2\rangle_c=t^{-2}\langle{\mathcal W}^2\rangle_c}\nonumber\\
&=&2t^{-1}\frac{(\det D+\hat{q}^2)\mbox{Tr}F}
{(\det D)^2}\hat{q}^2~.
\end{eqnarray}

For the stability of the NESS, we assumed at the beginning that the matrices $D$ and $F$ are positive definite.
Thus, $\det D>0$ and $\mbox{Tr} F>0$, which guarantee the positivity of $\langle{\mathcal W}\rangle_c$ and $\langle{\mathcal W}^2\rangle_c$.  From Eq.~(\ref{gaussian}), it can be shown that $s=\det D/(\det D+\hat{q}^2)< 1$,
which indicates that
$P(\sigma)$ is more distributed than the Gaussian distribution for nonzero $\hat{q}$.
We observe that all higher-order cumulants are also of order $t$,
i.e.~$\langle{\mathcal W}^n\rangle_c\sim t$, so $\langle\sigma^n\rangle_c\sim t^{-(n-1)}$.

We have performed numerical analysis to confirm our analytic results and
gain more insights on the work distribution function $P({\mathcal W})$.
Here we present numerical data taken at
$(k_1,k_2) = (4,1)$, $(\kappa_1,\kappa_2) = (2,1)$,
$(\alpha,\gamma) = (1,1)$, and $\epsilon = \sin\theta$ with $\theta=0.1$.
This system has a NESS with a nonzero value of $\hat{q}$.
We expect that the results do not depend on a specific
choice of parameter values as far as $\hat{q}$ is nonzero.

For $G_s$ and $G_a$, we adopt the AS choice given in
Eq.~(\ref{antisymmetric-gauge}) for convenience. Hence, the system is assumed to have an
initial probability distribution
$$P(q(0)) = |\det(2\pi A^{-1}(0))|^{-1/2} e^{-\frac{1}{2} q^T(0) \cdot A(0)
\cdot q(0)}$$
with $A(0) = \bar{G}_s$ and the work for a path $q(\tau)$ is
obtained using Eq.~(\ref{barW}).

The generating function ${\mathcal G}(\lambda)$ defined in Eq.~(\ref{G_lambda}) can be
estimated by direct numerical integrations of the Langevin equation in
Eq.~(\ref{langevin}). One starts from an initial state $q(0)$
drawn from the initial distribution $P(q(0))$.
The Langevin equation is then integrated with discretized time intervals
$$
q(\tau + \Delta t) = q(\tau) - F \cdot q(\tau) \Delta t + \Delta W(\tau)
$$
where $\Delta W(\tau) = (\Delta W_x(\tau),\Delta W_y(\tau))^T$ are correlated
random variables constructed as
$$
\left(\begin{array}{cc} \Delta W_x(\tau) \\
                        \Delta W_y(\tau) \end{array} \right) =
\sqrt{2\Delta t}\left(\begin{array}{cc} 1 & 0 \\
                        \sin\theta & \cos\theta \end{array} \right)
\left(\begin{array}{cc} \eta_x(\tau) \\
                        \eta_y(\tau) \end{array} \right) ~.
$$
Here, $\eta_{x,y}(\tau)$ are independent and identically distributed
Gaussian random variables with zero mean and unit variance. One can check
easily that such random variables $\Delta W(\tau)$ satisfy the required
correlation property $\langle \Delta W(\tau) \Delta W(\tau')^T\rangle = 2(\Delta t) D
\delta(\tau-\tau')$.

The work production is estimated as
$${\mathcal W}(\tau+\Delta t) = {\mathcal W}(\tau) - (q(\tau+\Delta
t)-q(\tau))^T \cdot \bar{G}_a \cdot q(\tau)~.$$
Repeating $N_S$ independent simulations, one obtains a numerical estimate
\begin{equation}
{\mathcal G}_{N}(t;\lambda) = \frac{1}{N_S} \sum_{n=1}^{N_s} e^{-\lambda
{\mathcal W}_n (t)}~.
\end{equation}

We can also utilize our analytic expression for $\mathcal{G}$ in Eq.~(\ref{lnG})
in order to get a more precise numerical estimate.
We first solve the NLDE for
$\tilde{A}(t;\lambda)$ in Eq.~(\ref{diff-eq3})
with the initial condition $\tilde{A}(0) = \bar{G}_s$.
The solution is obtained numerically in discretized times using the
recursion relation $\tilde{A}(\tau+\Delta t) = \tilde{A}(\tau) + \Delta t
\frac{d\tilde{A}(\tau)}{d\tau}$. The generating function $\mathcal{G}$
can be evaluated easily using the numerical solution $\tilde{A}$. The
generating function evaluated numerically using the analytic expression will
be denoted as $\mathcal{G}_A$.

\begin{figure}[th]
\includegraphics*[width=\columnwidth]{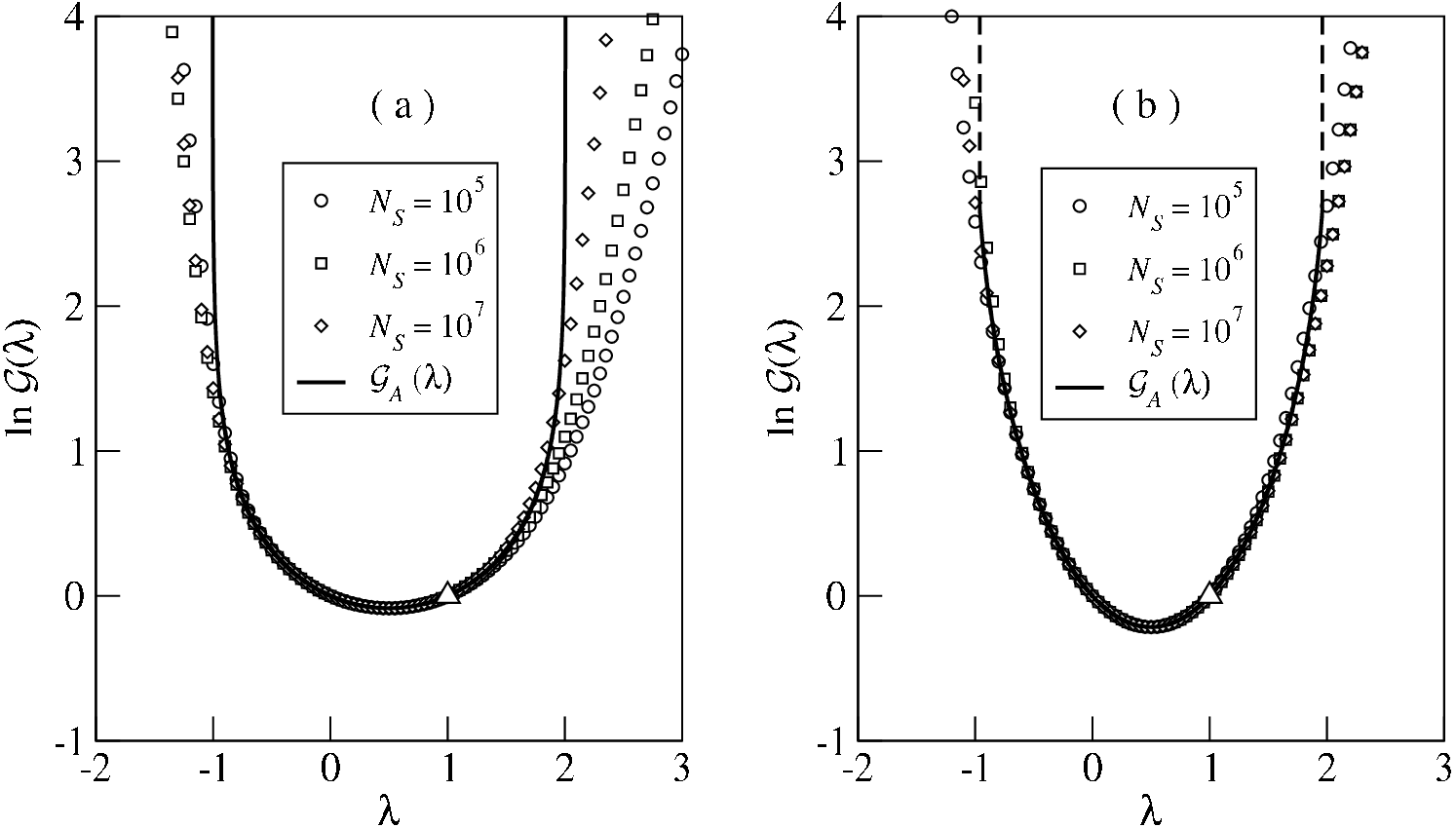}
\caption{Comparison of ${\mathcal G}_N(\lambda)$ and ${\mathcal
G}_A(\lambda)$ at $t=0.5$ in (a) and $t=2.0$ in (b). Symbols represent
${\mathcal G}_N$ and lines represent ${\mathcal G}_A$. Location of
the symbol $\triangle$ represents the Jarzynski equality $\mathcal{G}(1)=1$.
In (a),
${\mathcal G}_A$ diverges continuously as $\lambda$ approaches a threshold.
On the other hand, in (b), ${\mathcal G}_A$ remains finite up to a threshold
and diverges discontinuously beyond it.}\label{fig1}
\end{figure}

In Fig.~\ref{fig1}, we compare the results ${\mathcal G}_N(t;\lambda)$ and
${\mathcal G}_A(t;\lambda)$ at $t=0.5$ and $2.0$ obtained with
$\Delta t = 0.0001$ and $N_S \le 10^7$.
The two methods yield
almost identical results for small values of $|\lambda|$.
Both data confirm the Jarzynski equality ${\mathcal G}(\lambda=1)=1$ and the
Crooks fluctuation theorem $\mathcal{G}(\lambda) = \mathcal{G}(1-\lambda)$.
However, there is a
noticeable discrepancy at larger values of $|\lambda|$. Even the Crooks fluctuation
theorem seems to be violated in the ${\mathcal G}_N$ data.
One might suspect
a finite $\Delta t$ as a source of systematic errors. We have
also taken data with $\Delta t = 0.01, 0.001$, and $0.0001$ and found
no significant difference, which means that $\Delta t = 0.0001$ is already
small enough. In fact, the discrepancy is due to limited sampling
in obtaining ${\mathcal G}_{N}$. When $|\lambda|$ is large, ${\mathcal
G}(\lambda)$ is dominated by rare events with large $|{\mathcal W}|$.
If one compares ${\mathcal G}_N$ obtained from $N_S=10^5, 10^6$,
and $10^7$ samples, there are strong fluctuations at large values of
$|\lambda|$. This means that the tail property of the work distribution
function $P({\mathcal W})$ cannot be accessed from numerical simulations
even with $10^7$ samples.
On the contrary, the analytic formalism allows us to study the work
distribution function in detail without any statistics problem.

\begin{figure}
\includegraphics*[width=\columnwidth]{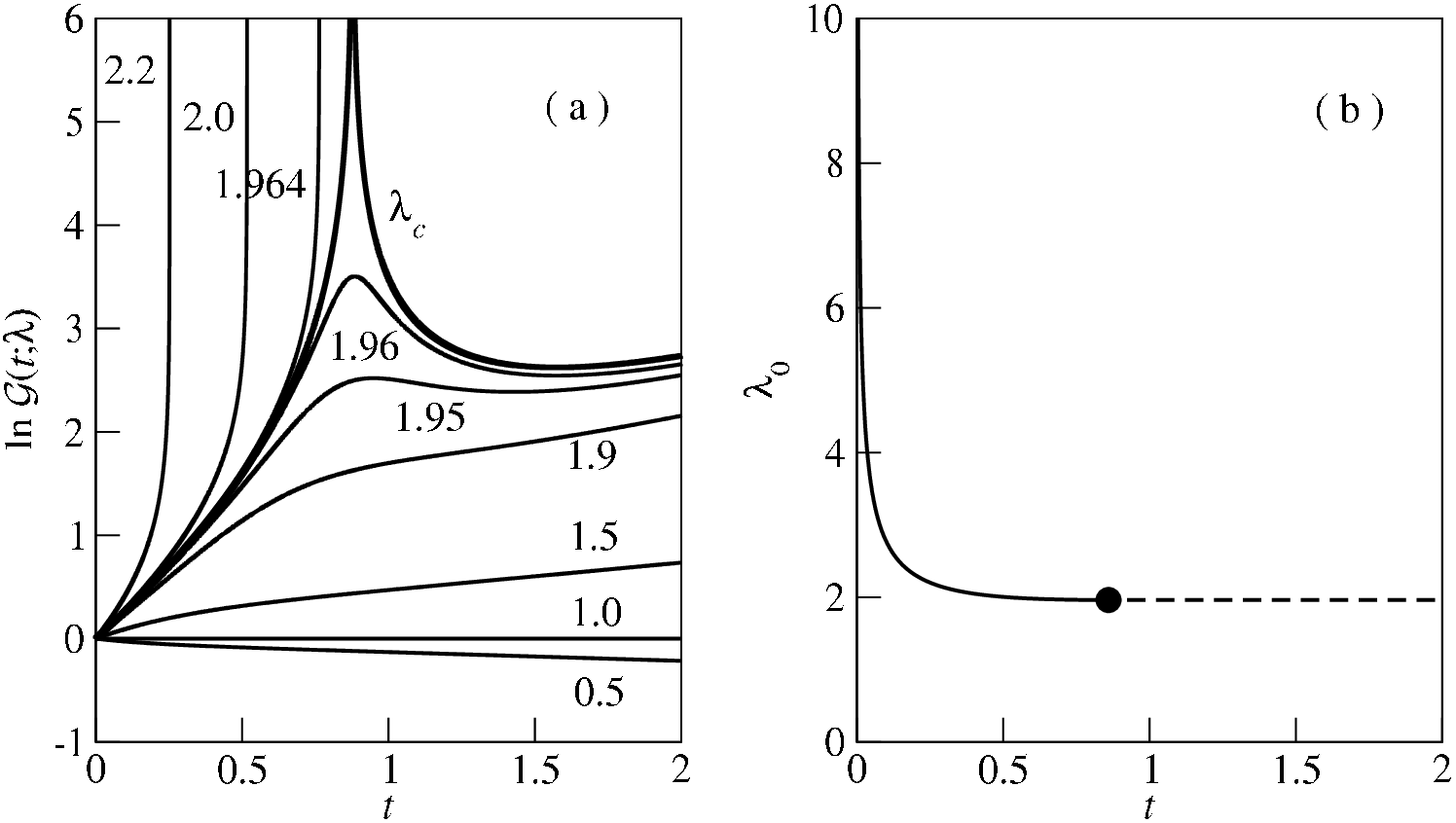}
\caption{(a) Plot of $\ln {\mathcal G}(t;\lambda)$ against $t$ at several
values of $\lambda$. (b) Threshold curve $\lambda_0(t)$ above which
${\mathcal G}(t;\lambda)$ is infinite. The symbol represents a point
$(t_c,\lambda_c) \simeq (0.86,1.962)$.}\label{fig2}
\end{figure}

Figure~\ref{fig1} shows that ${\mathcal G}(t;\lambda)$
becomes singular at $\lambda = \lambda_0$ and $1-\lambda_0$ with a
$t$-dependent threshold $\lambda_0=\lambda_0(t)>1$. The singular behavior is
evident in Fig.~\ref{fig2}(a), where we plot $\ln{\mathcal G}(t;\lambda)$ as a
function of $t$ at several values of $\lambda\geq 1/2$.
It suffices to consider $\lambda\geq 1/2$ because of the symmetry
${\mathcal G}(\lambda) = {\mathcal G}(1-\lambda)$.
When $\lambda < \lambda_c \simeq 1.962(1)$,
${\mathcal G}(t;\lambda)$ remains finite for all $t$.
On the other hand, it diverges at $t=t_c \simeq 0.86(1)$ at
$\lambda = \lambda_c$, and diverges at $t<t_c$ when $\lambda>\lambda_c$.
From these plots, we conclude that ${\mathcal G}(t;\lambda)$,
being viewed as a function of $\lambda$, diverges at $t$-dependent thresholds
$\lambda = \lambda_0(t)$ and $\lambda = 1-\lambda_0(t)$.
The threshold, numerically determined, is drawn in Fig.~\ref{fig2}(b).
Figure~2 also allows us to conclude that ${\mathcal G}(t;\lambda)$ diverges
continuously as $\lambda \to \lambda_0(t)$ ~(see the solid line in
Fig.~\ref{fig2}(b)) when $t \le t_c$. On the other hand, when $t>t_c$,
${\mathcal G}(t;\lambda)$ displays a discontinuous jump to infinity at
$\lambda=\lambda_c$~(see the dashed line in Fig.~\ref{fig2}(b)).
Interestingly, Fig.~\ref{fig2}(b) resembles a phase diagram of a system
having a tricritical point where a continuous phase transition line turns
into a discontinuous phase transition line.

\begin{figure}
\includegraphics*[width=\columnwidth]{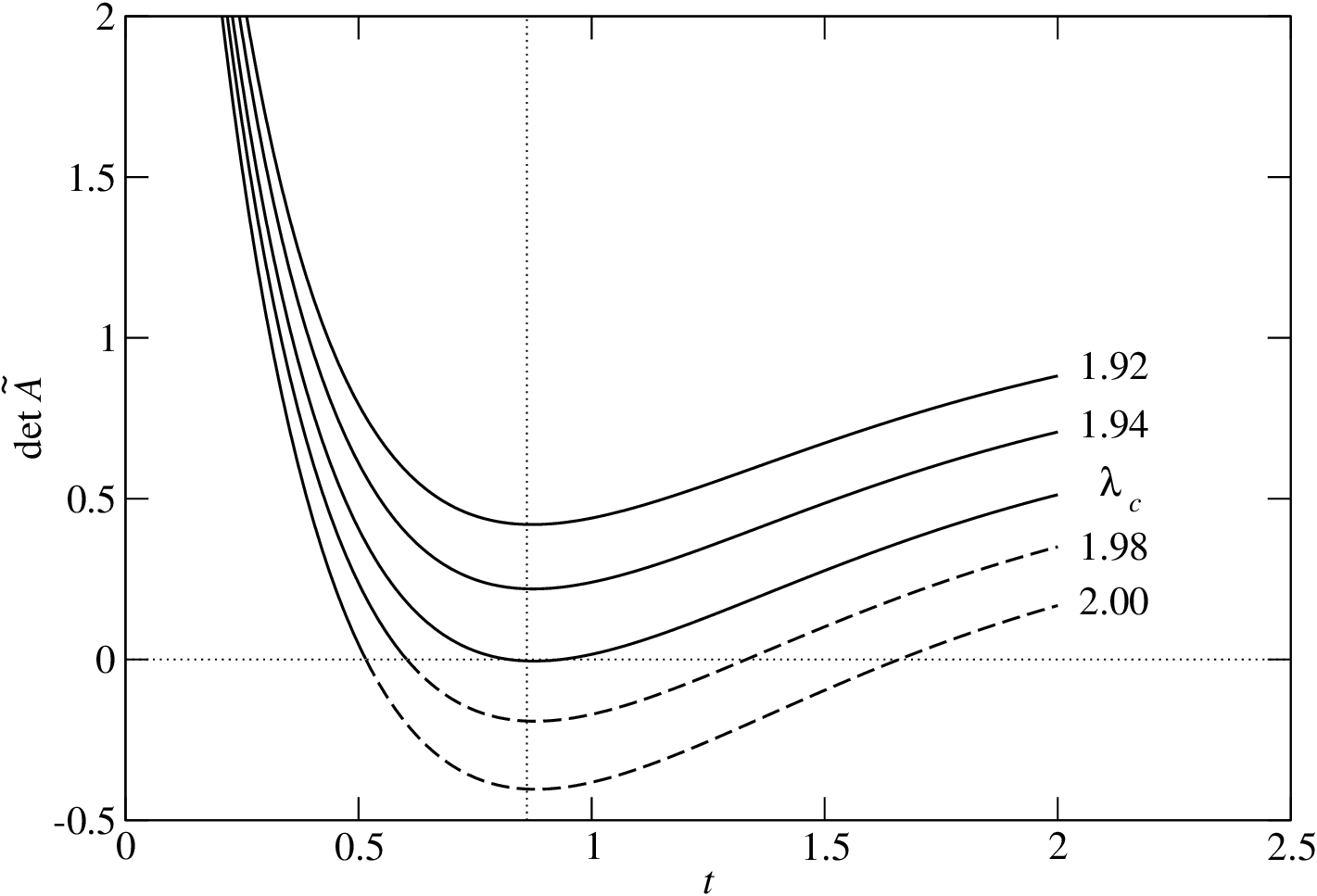}
\caption{Time evolution of $\det \tilde{A}(t;\lambda)$ near $t=t_c$ and $\lambda=\lambda_c$.
When $\lambda<\lambda_c\simeq 1.962$, the determinant seems to be always positive for any $t$.
In contrast, it becomes negative at finite $t$ for $\lambda>\lambda_c$
and the time evolution afterward (the dashed line) is
meaningless. When $\lambda=\lambda_c$, it is tangential to the $x$ axis
(the horizontal dotted line) at $t=t_c$ (the vertical dotted line).}\label{fig3}
\end{figure}

Origin and nature of the divergence are understood from the analytic
expression for $\ln \mathcal{G}$ in Eq.~(\ref{lnG}). Due to the logarithmic
boundary term, $\mathcal{G}(t;\lambda)$ is well-defined only when
$\det (\tilde{A}(t';\lambda)+\lambda \delta G_s)$ is positive for all $t'<t$.
In contrast, there is no singularity in the bulk term for any $t$.
From the numerical solution of the NLDE, Eq.~(\ref{diff-eq3}),
we observed that $\det(\tilde{A}(t;\lambda))$~($\delta G_s=0$ with
the AS choice) behaves as
\begin{equation}
\det(\tilde{A}(t;\lambda)) \simeq a(\lambda_c-\lambda) + b (t-t_c)^2
\end{equation}
near $\lambda = \lambda_c$ and $t=t_c$  with positive constants $a$ and $b$,
see Fig.~\ref{fig3}.
This behavior explains the singularity in $\mathcal{G}$.

When $t \le t_c$, the determinant becomes zero at $\lambda=\lambda_0$ with
\begin{equation}
\lambda_0 \simeq \lambda_c + \frac{b}{a}(t-t_c)^2 ~.
\end{equation}
Consequently, $\mathcal{G}$ diverges {\em continuously} as
\begin{equation}\label{Gsingularity1}
\mathcal{G}(t;\lambda) \sim (\lambda_0-\lambda)^{-1/2}
\end{equation}
as $\lambda$ approaches $\lambda_0$ from below.

The determinant is always positive when $\lambda<\lambda_c$, while it
becomes negative at $t<t_c$ when $\lambda>\lambda_c$. Hence, when $t>t_c$
$\mathcal{G}(t;\lambda)$ remains finite up to $\lambda<\lambda_c$ and
diverges {\em discontinuously} at $\lambda=\lambda_c$. As $\lambda$
approaches $\lambda_c$ from below, it behaves regularly as
\begin{equation}\label{Gsingularity2}
\mathcal{G}(t;\lambda) \sim h +a(\lambda_c - \lambda)^1
\end{equation}
with the $t$-dependent constant $h$.

The singularities in ${\mathcal G}(\lambda)=\int d{\mathcal W} P({\mathcal W})e^{-\lambda {\mathcal W}}$
at $\lambda=\lambda_0$ and $1-\lambda_0$
indicate that the PDF  $P(\mathcal{W})$ has exponential tails
\begin{equation}
P({\mathcal W}) \sim \left\{
\begin{array}{ll}
{\mathcal W}^{-r}~ e^{-\mathcal{W}/\mathcal{W}_+} &
\mbox{ for } {\mathcal W}\to \infty \\ [3mm]
(-{\mathcal W})^{-r}~ e^{-{\mathcal W}/\mathcal{W}_-} &
\mbox{ for } {\mathcal W}\to -\infty
\end{array}\right.
\label{pdf_tail}
\end{equation}
with characteristic works $\mathcal{W}_+$ $(>0)$ and $\mathcal{W}_-$ $(<0)$, and possible power-law
corrections with exponent $r$. The power-law prefactor is necessary in order
to account for the way how $\mathcal{G}$ becomes singular at
$\lambda=\lambda_0$. From the Crooks fluctuation theorem,
$P(\mathcal{W}) = e^{\mathcal{W}} P(-\mathcal{W})$,
the exponent $r$ should be the same for both tails, and it suffices to
consider one of the tails.

It is easy to check that the negative
tail yields $\mathcal{G}(\lambda) \sim (1/|\mathcal{W}_-| - \lambda)^{r
-1}$. Comparing it with Eqs.~(\ref{Gsingularity1}) and
(\ref{Gsingularity2}), we find that the characteristic work is given by
$\mathcal{W}_- = -1/\lambda_0$ and that the exponent is given by
\begin{equation}
r = \left\{ \begin{array}{ccc}
\frac{1}{2} & \mbox{for } t\leq t_c ~, \\ [3mm]
2           & \mbox{for } t > t_c ~.
\end{array} \right.
\label{exponent}
\end{equation}
The positive tail has $\mathcal{W}_+ = 1/(\lambda_0 -1)$ and the same
exponent $r$.

\begin{figure}
\includegraphics*[width=\columnwidth]{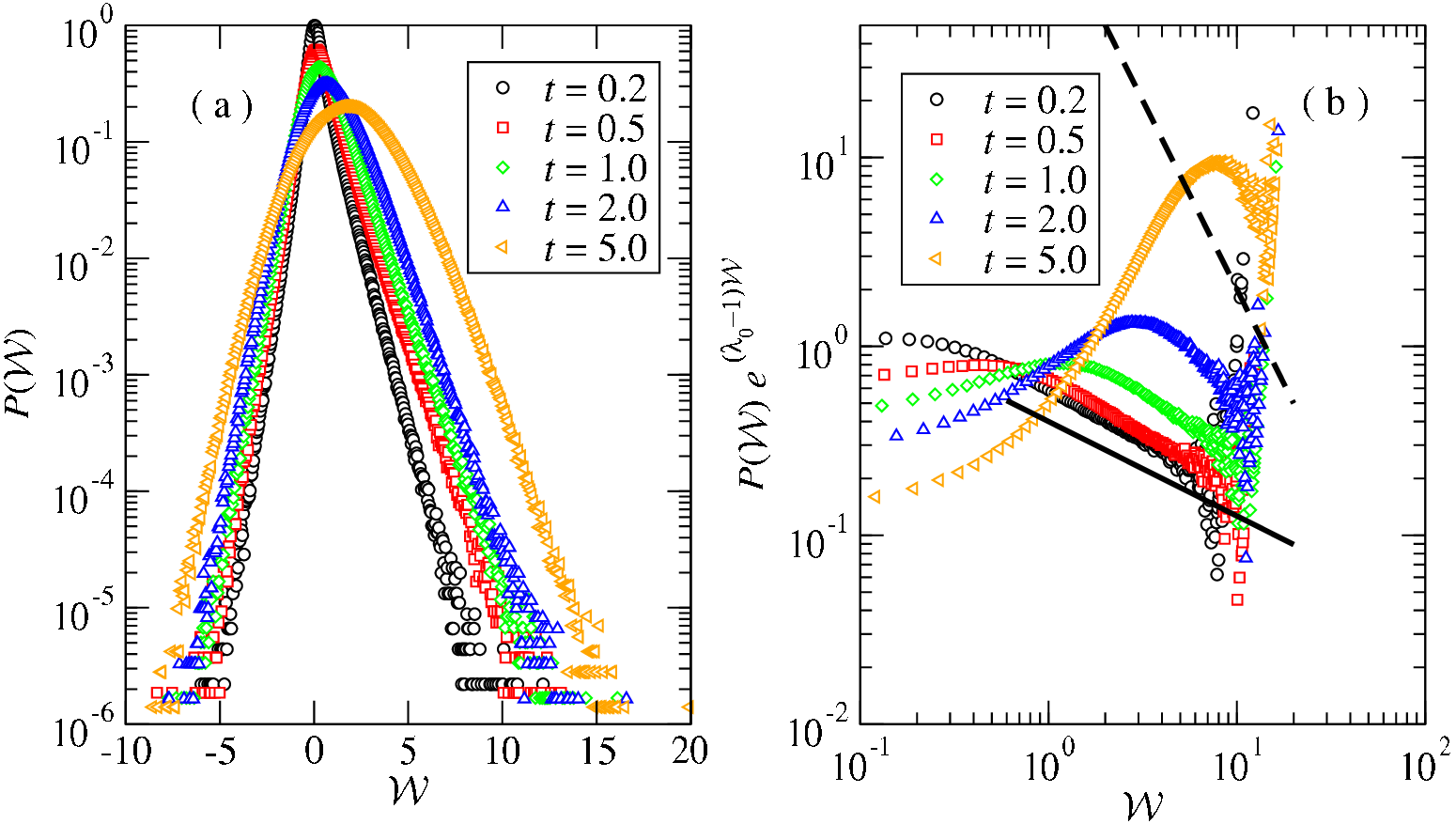}
\caption{(Color online) (a) Plot of $P(\mathcal{W})$ versus $\mathcal{W}$ at several
values of $t$.
(b) Plot of $P(\mathcal{W}) e^{(\lambda_0-1)\mathcal{W}}$ versus
$\mathcal{W}$ for $\mathcal{W}>0$. The solid and dashed lines are guides to eyes with slope
$-1/2$ and $-2$, respectively.}\label{fig4}
\end{figure}

Numerical data are consistent with the tail property in
Eq.~(\ref{pdf_tail}). The PDF $P(\mathcal{W})$ was
obtained from numerical simulations of $N_S = 10^7$ samples.
Figure~\ref{fig4}(a) presents the plot of $P(\mathcal{W})$ against
$\mathcal{W}$ at several values of $t$ in the semi-log scale.
As expected, $P(\mathcal{W})$ becomes more distributed and the mean work production
$\langle\mathcal{W}\rangle$ increases with time $t$. Moreover,
the exponential tails are clearly seen in all plots with characteristic works
(slopes of plots in Fig.~\ref{fig4}(a)) saturating with time $t$.

We also present the log-log plot of $P(\mathcal{W})e^{(\lambda_0-1) \mathcal{W}}$ against
$\mathcal{W}$ for $W>0$ in Fig.~\ref{fig4}(b). We use the threshold value $\lambda_0 (t)$
obtained from the singularities in ${\mathcal G}_A$, which are shown in Fig.~\ref{fig2}(b).
One can see that the $P(\mathcal{W})e^{(\lambda_0-1) \mathcal{W}}$ has a
power-law tail as predicted in Eq.~(\ref{pdf_tail}).
When $t=0.2 \mbox{ and } 0.5~ (< t_c)$, the power-law tail is manifest and the
exponent is in good agreement with the analytic prediction $r=1/2$. When
$t=2.0 \mbox{ and } 5.0~(>t_c)$, the power-law sets in at larger values of
$\mathcal{W}$ and the exponent value is drifting with increasing $\mathcal{W}$.
In this case, we need much more samples ($N_S\gg 10^7$) to get good statistics for rare events at large $\mathcal{W}$,
in order to extract reliable quantitative information on the power-law tail.
Nevertheless, one can see that the exponent value becomes close to $r=2$ for large $t$.
At $t=1.0$, the crossover effect dominates the numerical data,
since $t$ is too close to $t_c \simeq 0.86$.

It is not surprising to see the exponential tail with a power-law prefactor with the
exponent $r=1/2$.  This exponential tail has been also observed in other NEQ systems~\cite{comment,chatterjee,crooks-jarzynski,kwon3}.
When the PDF is given as $P(q)\sim e^{-\Phi(q)}$ with
the energy-like function $\Phi(q)$ quadratic in $q$ as in our case, the PDF for any quantity ${\mathcal A}$
also quadratic in $q$ should become $P({\mathcal A})\sim {\mathcal A}^{-1/2} e^{-{\mathcal A}/{\mathcal A}_c}$ with
the characteristic value of ${\mathcal A}_c$. As the work ${\mathcal W}$ is simply quadratic in $q$ and proportional to $t$,
at least for a short time $t$, i.e.~${\mathcal W}\sim t q^2$ as in our case, our finding in Eqs.~(\ref{pdf_tail}) and (\ref{exponent}) can be understood with ${\mathcal A}_c$ increasing in time $t$.

However, there is a sharp dynamic phase transition at $t=t_c$ beyond which ${\mathcal A}_c$ is constant of time and the
power-law exponent changes from $1/2$ to $2$. This implies that the characteristic positive work production
${\mathcal W}_+=1/(\lambda_0(t)-1)$  increases monotonically in time for $t<t_c$,
but saturates to a finite value $1/(\lambda_c-1)$ at $t=t_c$. For $t>t_c$, the characteristic work
production remains unchanged and only the power-law prefactor adjusts the PDF accordingly.
We do not have an intuitive understanding for the dynamic phase transition at this moment, except that
the transition time $t_c$ may be related to the intrinsic relaxation time of the system.
It calls for a further study to understand this phase transition, which is currently under investigation.

\section{Summary and Discussion}

We have studied NEQ fluctuations for high-dimensional diffusion dynamics driven
by a linear drift force. The drift force is not derivable from a scalar potential function
and the noises are not identical white noises in components with possible correlations.
In general, these NEQ features generate a circulating probability current at the steady state,
with the drift force inward to the origin (the force matrix $F$ is positive definite).
It is interesting to notice that the equilibrium can be restored with a specific combination
of these two NEQ features by satisfying the detailed balance condition ($DF=F^TD$).

Recent experiments on an optical trap report that particles can be confined in a field-induced
potential well, approximated by an asymmetric harmonic potential. It may be possible to
apply our study to such experiments. However, there are some technical issues to be resolved in experiments.
In particular, in order to break the detailed balance, the noises should be applied
in such a way that their principal axes do not coincide with those of the potential well~\cite{filliger}.

In this paper, we started with the Langevin equation with an arbitrary drift force and
arbitrary additive noises in high dimensions without any thermodynamic origin. The generalized thermodynamic
quantities like energy, work, and heat are defined, with which the Crooks fluctuation theorems are reproduced.
For the case of the linear drift force, we derive the exact evolution function of the PDF analytically.
More importantly, we analyzed the work PDF, $P({\mathcal W})$ through the generating function method, which showed
an interesting dynamic phase transition in the exponential tail shape of $P({\mathcal W})$.
As the tail property is governed by rare events, it is crucial to develop an analytic theory to show this subtle
dynamic phase transition, which can be hardly identified by numerical simulations only.
Even though we expect that this phase transition is one of the generic features found in the NEQ systems
described by Langevin equations, it would be a big challenge to understand analytically how and when this can arise.

Finally, in the case of nonlinear forces,
it has been found that there is an additional current which shifts the probability maximum from the fixed point
at the origin ~\cite{kwon2}. It may have the same origin as the noise-driven directed current in extended systems, \cite{prost,doering}.
The ingredients for those currents are: (a) non-linearity of the inward drift force and high-dimensional noises for the confined system (b) asymmetric (ratchet-type) potential and an additional noise for the extended system.
The NEQ fluctuation theory for this non-linear diffusion dynamics will be another challenging topic.

\begin{acknowledgments}
We would like to thank David Thouless, Marcel den Nijs, Hyungtae Kook, Su-Chan Park, and Kyung Hyuk Kim for helpful comments. We also thank Ping Ao for stimulating discussions and Hong Qian for introducing his enlarged view on a NEQ principle.
We thank Korea Institute for Advanced Study for providing computing resources
(KIAS Center for Advanced Computation Linux Cluster System) for this work.
CK greatly appreciates the Condensed Matter Group at University of Washington  for giving him a warm hospitality and good opportunity to use full academic facilities during his sabbatical year.
This work was supported by Mid-career Researcher Program through
NRF grant (No.~2010-0026627) funded by the MEST.
\end{acknowledgments}

\appendix
\section{Time-dependent PDF}
\label{path-integration}
We keep the source field for later use to compute the generating functional. The initial probability distribution is chosen as
\begin{equation}
P(q(0))={\mathcal N}_0e^{-\frac{1}{2}q^{T}(0)\cdot A(0)\cdot q(0)}~,
\end{equation}
with ${\mathcal N}_0=|\det(2\pi A^{-1}(0))|^{-1/2}$ and $A^T(0)=A(0)$. We consider the sequence of discrete times, $0=t_0<t_1<t_2<\cdots <t_{N}=t$ with time interval $\Delta t=t/N$ in the $N\to\infty$ limit. Denoting $q_i\equiv q(t_i)$ with $q_0=q(0)$ and $q_N=q(t)=q$, and $A_0=A(0)$, we write $P(q,t;l)$ in the pre-point representation
(set $q(\tau)=q_{i-1}$ for $t_{i-1}\le \tau\le t_i$)
as
\begin{eqnarray}
\lefteqn{P(q,t;l)={\mathcal N}_0\int \prod_{i=0}^{N-1} \frac{dq_i}{|\det(4\pi\Delta t D)|^{1/2}}}\nonumber\\
&&\times e^{-\frac{1}{4\Delta t}\sum_{i=1}^{N}\left(q_{i}-q_{i-1}+\Delta t F\cdot q_{i-1}\right)^{T}\cdot
D^{-1}\cdot \left(q_{i}-q_{i-1}+\Delta t F\cdot q_{i-1}\right)}  \nonumber\\
&&\times e^{\Delta t\sum_{i=0}^{N}l_i^{T}\cdot q_{i}-\frac{1}{2}q_0^{T}\cdot A_0 \cdot q_0} \nonumber\\
&=&{\mathcal N}_0\int\prod_{i=0}^{N-1} \frac{dq_i}{|\det(4\pi\Delta t D)|^{1/2}}
\left[\prod_{i=1}^{N}T_{i,i-1}\right]e^{-\frac{1}{2}q_0^T\cdot A_0\cdot q_0}~.\nonumber\\
&&
\end{eqnarray}
The transfer matrix $T_{i,i-1}$ is given as
\begin{eqnarray}
\ln T_{i,i-1}(l_{i-1})&&=
-\frac{1}{4\Delta t} \left[ \left(q_i-Vq_{i-1}\right)^T\cdot D^{-1}\cdot \left(q_i-Vq_{i-1}\right)\right]\nonumber\\
&& ~~~~~~+ (\Delta t) l^T_{i-1}\cdot q_{i-1},
\end{eqnarray}
where
\begin{equation}
V=1-\Delta t F.
\end{equation}

Integrating over $q_0$, we obtain
\begin{eqnarray}
I_0&=&\int\frac{dq_0}{|\det(4\pi\Delta t D)|^{1/2}} T_{1,0}e^{-\frac{1}{2}q_0^T\cdot A_0\cdot q_0} \nonumber\\
&=& |\det(DB_0)|^{-1/2}e^{-\frac{1}{2}q_1^T\cdot A_1\cdot q_1}\nonumber\\
&&\times e^{(\Delta t)^3 l_0^T\cdot B_0^{-1}\cdot l_0+(\Delta t) l_0^T\cdot B_0^{-1}V^TD^{-1}\cdot q_1}~,
\end{eqnarray}
where
\begin{eqnarray}
A_1&=&\frac{1}{2\Delta t}\left(D^{-1}-D^{-1}VB_0^{-1}V^{T}D^{-1}\right)\\
B_0&=&V^{T}D^{-1}V+2\Delta t A_0~,
\end{eqnarray}
where $A_1$ and $B_0$ are both symmetric.
Then we can find
\begin{eqnarray}
\lefteqn{T_{2,1}(l_1)I_0={T}_{2,1}(\tilde{l}_1)e^{-\frac{1}{2}q_1^T\cdot A_1\cdot q_1}}
\nonumber\\
&&\times |\det(DB_0)|^{-1/2}e^{(\Delta t)^3 l_0^T\cdot B_0^{-1}\cdot l_0}~.
\end{eqnarray}
where $\tilde{l}_1=l_1+D^{-1}VB_0^{-1}\cdot l_0$. Notice that the integration of $T_{1,0}$ over $q_0$ results in a change $\tilde{l}_1$ in the source field
of $T_{2,1}$. It happens iteratively for subsequent integrations over the rest of $q_{i}$. We show it explicitly for the next step,
\begin{eqnarray}
I_1&=&\int\frac{dq_1}{|\det(4\pi\Delta t D)|^{1/2}}
T_{2,1}(\tilde{l}_1)
e^{-\frac{1}{2}q_1^T\cdot A_1\cdot q_1} \nonumber\\
&=& |\det(DB_1)|^{-1/2}e^{-\frac{1}{2}q_2^T\cdot A_2\cdot q_2}\nonumber\\
&&\times e^{(\Delta t)^3 \tilde{l}_1^T\cdot B_1^{-1}\cdot \tilde{l}_1 +
(\Delta t) \tilde{l}_1^T\cdot B_1^{-1}V^TD^{-1}\cdot q_1}~,
\end{eqnarray}
where
\begin{eqnarray}
A_2&=&\frac{1}{2\Delta t}\left(D^{-1}-D^{-1}VB_1^{-1}V^{T}D^{-1}\right)\\
B_1&=&V^{T}D^{-1}V+2\Delta t A_1~.
\end{eqnarray}
Then we get
\begin{eqnarray}
\lefteqn{T_{3,2}(l_2)I_1={T}_{3,2}(\tilde{l}_2)e^{-\frac{1}{2}q_1^T\cdot A_2\cdot q_1}}
\nonumber\\
&&\times |\det(DB_1)|^{-1/2}e^{(\Delta t)^3 \tilde{l}_1^T\cdot B_1^{-1}\cdot \tilde{l}_1}~.
\end{eqnarray}
where $\tilde{l}_2=l_2+D^{-1}VB_1^{-1}\cdot\tilde{l}_1$.
We repeat the integrations over all $q_i$ up to $i=N-1$. Finally, we reach the result:
\begin{eqnarray}
\lefteqn{P(q_N, t_N;l)=|\det(2\pi A_0^{-1})|^{-1/2}\prod_{i=0}^{N-1}|\det(DB_i)|^{-1/2}}\nonumber\\
&&\times e^{-\frac{1}{2}q_N^T\cdot A_N\cdot q_N+(\Delta t) \tilde{l}_N^T\cdot q_N+(\Delta t)^3 \sum_{i=0}^{N-1}\tilde{l}_i^T\cdot B_i^{-1}\cdot \tilde{l}_i}~.
\label{time-dep-prob1}
\end{eqnarray}
We have the recursion relations:
\begin{eqnarray}
A_i&=&\frac{1}{2\Delta t}\Big(D^{-1}-D^{-1}VB_{i-1}^{-1}V^{T}D^{-1}\Big)\\
B_{i-1}&=&V^{T}D^{-1}V+2\Delta t A_{i-1}\label{Bi}\\
\tilde{l}_i&=&l_i+D^{-1}VB_{i-1}\cdot\tilde{l}_{i-1},\label{li}
\end{eqnarray}
for $i=1,\ldots, N$ with $\tilde{l}_0=l_0$.

$A_i$ determines the intermediate probability distribution function at time $t_i$. We can find a simple recursion relation for $A_i^{-1}$. First we observe the following relation:
\begin{eqnarray}
A_i&=&\frac{1}{2\Delta t}\left\{D^{-1}-\Big(D+2\Delta t D(V^T)^{-1}A_{i-1}V^{-1}D\Big)^{-1}\right\}\nonumber\\
&=&(V^T)^{-1}A_{i-1}V^{-1}D\Big(D+2\Delta t D(V^T)^{-1}A_{i-1}V^{-1}D\Big)^{-1}\nonumber\\
&=&\Big(2\Delta t D+VA_{i-1}^{-1}V^T\Big)^{-1}~.
\end{eqnarray}
Therefore we get
\begin{equation}
A_i^{-1}=2\Delta t D+VA_{i-1}^{-1}V^T~,
\label{recursion1}
\end{equation}
which leads to
\begin{eqnarray}
A_i^{-1}&=&2\Delta t D +2\Delta t VDV^T +V^2A_{i-2}^{-1}(V^T)^2 \nonumber\\
&=&2\Delta t\sum_{j=0}^{i-1}V^{j}D(V^T)^j+V^iA_0^{-1}(V^T)^i~.
\end{eqnarray}
In the continuum limit, we have
\begin{equation}
A^{-1}(t)=2\int_{0}^t d\tau e^{-F\tau}De^{-F^T \tau}+e^{-Ft}A^{-1}(0)e^{-F^T t}~.
\end{equation}

We use the integral identity as
\begin{eqnarray}
\lefteqn{\int_0^t d\tau e^{-H\tau}Ce^{-H^T\tau}=}\nonumber\\
&&~~~~~\frac{1}{2}H^{-1}\left[(C+E)- e^{-Ht}(C+E)e^{-H^Tt}\right],
\label{integral}
\end{eqnarray}
for arbitrary matrices $C$ and $H$ where
$E$ is determined from the equation
\begin{equation}
HE+EH^T=HC-CH^T~.
\end{equation}
Note that $E$ is antisymmetric (symmetric) if $C$ is symmetric (antisymmetric).

Here, with $C=D$ and $H=F$, the antisymmetric matrix $E$ becomes equal to $Q$ from Eq.~(\ref{anti-symmetric-matrix}).
Then $H^{-1}(C+E)=F^{-1}(D+Q)=U^{-1}$ from Eq.~(\ref{U}). Therefore, we have
\begin{equation}
A^{-1}(t)=U^{-1} + e^{-Ft}(A^{-1}(0)-U^{-1})e^{-F^T t}~.
\label{time-dep-variance1}
\end{equation}

Using Eq.~(\ref{recursion1}), the recursion relation for $B_{i-1}$, Eq.~(\ref{Bi}), becomes
\begin{eqnarray}
B_{i-1}&=&A_{i-1}V^{-1}\Big(VA_{i-1}^{-1}V^T+2\Delta t D\Big)D^{-1}V
\nonumber\\
&=&A_{i-1}V^{-1}A_i^{-1}D^{-1}V~.
\label{eq-B}
\end{eqnarray}
Therefore we get
\begin{equation}
\det(DB_{i-1})=\frac{\det A_{i-1}}{\det A_i}~,
\end{equation}
which simplifies the normalization factor in Eq.~(\ref{time-dep-prob1}) as
\begin{eqnarray}
\lefteqn{|\det(2\pi A_0^{-1})|^{-1/2}\prod_{i=0}^{N-1}|\det(DB_i)|^{-1/2}}\nonumber\\
&=&\left|\det(2\pi A_0^{-1})\frac{\det A_0}{\det A_1}\frac{\det A_1}{\det A_2}\cdots\frac{\det A_{N-1}}{\det A_N}\right|^{-1/2}\nonumber\\
&=&|\det(2\pi A_N^{-1})|^{-1/2}~.
\end{eqnarray}
It is the correct normalization factor for the final time-dependent probability distribution, $P(q,t)=P(q,t;l=0)$:
\begin{equation}
P(q,t)=|\det(2\pi A^{-1}(t))|^{-1/2}e^{-\frac{1}{2}q^T\cdot A(t)\cdot q}~.
\label{time-dep-prob2}
\end{equation}

\section{Generating functional}
\label{generating-functional}
The generating functional $Z[l]=\int dq P(q,t;l)$ is obtained by integrating out Eq.~(\ref{time-dep-prob1}) over $q_N$:
\begin{eqnarray}
Z[l]&=&e^{(\Delta t)^3\sum_{i=0}^{N}\tilde{l}_i^T\cdot B_i^{-1}\cdot\tilde{l}_i}\nonumber\\
&\equiv& e^{\frac{1}{2}(\Delta t)^2\sum_{i,j} l_i^T\cdot \Gamma_{ij}\cdot l_j}~,
\end{eqnarray}
where $B_N\equiv 2(\Delta t)A_N$. Investigating the recursion relation for $\tilde{l}_i$, Eq.~(\ref{li}),
we first show that for $i>j$
\begin{eqnarray}
\Gamma_{ii}&=&2\Delta t B_i^{-1}+C_i^T \Gamma_{i+1,i+1} C_i \\
\Gamma_{ij}&=&\Gamma_{ii} C_{i-1}C_{i-2}\cdots C_j
\end{eqnarray}
where
\begin{equation}
C_i=D^{-1}VB_i^{-1}=A_{i+1} V A_i^{-1}~.
\end{equation}
Using $\Gamma_{NN}=A_N^{-1}$ and Eq.~(\ref{eq-B}),
\begin{eqnarray}
\Gamma_{N-1,N-1}&=&2\Delta t V^{-1}D A_N V A_{N-1}^{-1}+C_{N-1}^TA_N^{-1}C_{N-1}\nonumber\\
&=&2\Delta t V^{-1}D A_N V A_{N-1}^{-1}+A_{N-1}^{-1}V^TA_{N}VA_{N-1}^{-1}\nonumber\\
&=&V^{-1}\Big(2\Delta t D+VA_{N-1}^{-1}V^T\Big)A_N V A_{N-1}^{-1}\nonumber\\
&=&A_{N-1}^{-1}~.
\end{eqnarray}
We can prove by induction
\begin{equation}
\Gamma_{ii}=A_i^{-1}~.
\label{diagonal}
\end{equation}
By noting
\[C_{i-1}C_{i-2}\cdots C_j=A_i V^{i-j} A_j^{-1}~,\]
we find
\begin{equation}
\Gamma_{ij}=V^{i-j}A_j^{-1}~. \label{off-diagonal}
\end{equation}
Note that for $j>i$, $\Gamma_{ij}=\Gamma_{ji}^T$. In the continuum limit, we finally get Eq.~(\ref{correlation}).


\begin{thebibliography}{99}
\bibitem{FP-derivation} This differential equation can be directly derived from the Fokker-Planck equation, Eq.~(\ref{fokker-planck}), by assuming the Gaussian form of the PDF as in Eq.~(\ref{kernel}).
\bibitem{midpoint}
For example, in equilibrium, one can check the energy-work relation such as
$\int_0^t d\tau \dot{q}\cdot\nabla\Phi=\Delta\Phi$ (see Eq.~(\ref{detailed-fluctuation-relation})),
which is correct only in the mid-point representation.
One finds an extra term in the other representation, which can be shown by expanding $\Phi(q_i)$ up to the order
of  $(q_i-q_{i-1})^2$ which is also ${\mathcal O}(\Delta t)$.
\end{thebibliography}
\end{document}